\documentstyle[11pt,epsfig]{article}
\newcommand{\bea}{\begin{eqnarray}}
\newcommand{\eea}{\end{eqnarray}}

\def\simlt{\stackrel{<}{{}_\sim}}
\def\simgt{\stackrel{>}{{}_\sim}}

\begin{document}
\begin{titlepage}

\thispagestyle{empty}

\vspace{0.2cm}

\title{ Dipole coefficients in $B\rightarrow X_{s} \gamma$ in supersymmetry 
with large $\tan\beta$ and explicit CP violation
\author{M\"{u}ge Boz$^{a}$ and Nam{\i}k K. Pak$^{b}$\\
$^a$ Hacettepe University, Department of Physics,\\  
06532 Ankara, Turkey\\
$^b$Middle East Technical University, Department of Physics,\\ 06531 Ankara,
Turkey\\[7mm]}}
\date{}
\maketitle

\begin{center}\begin{minipage}{5in}

\begin{center} ABSTRACT\end{center}
\baselineskip 0.2in
{We perform a detailed study of the electric
and chromoelectric dipole coefficients
in $B\rightarrow X_s \gamma$ decay in
a supersymmetric scheme  with explicit CP violation.
In our analysis, we adopt the  minimal flavor violation scheme by taking into account
the  $\tan\beta$-enhanced large contributions beyond the leading order.
We show that the coefficients can
deviate from the SM prediction significantly
in both real and imaginary directions.
Experimental bounds still allow for large deviations from the SM
predictions for both dipole coefficients such that the CP asymmetry is as
large as $\pm 8\%$. There are further implications of these coefficients
for the charmless hadronic and semileptonic $B$ decays. As a direct application of our analysis,
we have discussed  $\Lambda_b \rightarrow
\Lambda \gamma$ decay.}

\end{minipage}
\end{center}
\end{titlepage}

\eject
\rm
\baselineskip=0.25in

The Operator Product Expansion (OPE) combined with the heavy quark effective 
theory (HQET) forms the basic tool in analyzing the decays as well as
productions of hadrons (See the review \cite{ope}). Basically, the effective 
Hamiltonian describing the scattering processes can be expanded in a series of local operators
(whose hadronic matrix elements constitute the long--distance effects) 
with Wilsonian coefficients (which are generated by the 
short--distance effects). Among all the hadronic scattering processes the 
rare ones are particularly important,  as the contributions of the standard 
electroweak theory (SM) and those of the `new physics (NP)' arise 
at the same loop level thus suffering no relative loop suppressions. 
Furthermore, decays of the $b$--flavored hadrons ($B$, $B^{\star}$, $\Lambda_{b}$, $\cdots$),
compared to strange and charmed ones,  are especially important
as  for such systems the HQET is fully  applicable, and via the OPE,
one can both test the SM and search for possible NP effects, by confronting
the associated Wilson coefficients with the experiment. 

In this work we will analyze the electric and chromoelectric 
dipole coefficients (denoted hereafter by ${\cal{C}}_{7}$ and 
${\cal{C}}_{8}$, respectively) describing the short--distance
physics effects in rare $B$ decays ($e.g.$ $B\rightarrow X_s \gamma$,
$B\rightarrow K^{\star}(892) \gamma$ and $B\rightarrow X_{s} \ell^{+}
\ell^{-}$, $\cdots$) \cite{munz}, by taking into account 
existing experimental results. The electric dipole 
coefficient ${\cal{C}}_{7}$, rescaled to $\mu=m_{b}$ level, is 
directly constrained by the experimental result on 
$B\rightarrow X_s \gamma$ \cite{exp}. However, the 
situation for the  coefficient ${\cal{C}}_{8}$
is obscured by the fact that the gluonic decay
$b\rightarrow s g$ is not directly accesible in experiments.
However, this very coefficient plays an important 
role in the charmless hadronic $B$ decays. For example, the theoretical
predictions for the inclusive semileptonic decay rate 
$\Gamma(b\rightarrow c e \nu)$ and the charm multiplicity 
in $B$ meson decays are significantly higher than the
experimental results (See \cite{besmer} and references
therein). The  most plausable solution to this discrepancy stems from
possible enhancement of the chromoelectric coefficient 
${\cal{C}}_{8}$ by NP effects. Moreover, it is 
the relative phase between ${\cal{C}}_{8}$ and
${\cal{C}}_{7}$ which determines the CP asymmetry
in $B\rightarrow X_s \gamma$, which will be measured
in near--future $B$ factories \cite{neubert1}. Consequently,
it is essential to determine the size and phase of
the gluonic coefficient in regions of the
parameter space of NP where the theoretical predictions for $B\rightarrow X_s \gamma$
agrees with the experiment. 

In what follows, we will take  low--energy minimal 
supersymmetry (SUSY) with explicit CP violation
as the NP candidate. We will adopt the
minimal flavor violation scenario (MFV), take
into account the $\tan\beta$--enhanced large
contributions beyond the leading order (LO). The
inclusive mode $B\rightarrow X_s \gamma$ has 
already been analyzed within such a 
scheme by \cite{giudice} (in CP--conserving
SUSY), and has been furthered by \cite{olive} 
to the CP--violating SUSY concluding a
sizable CP asymmetry, which can compete 
with the experiment in near future. However,
in both \cite{giudice} and \cite{olive}
the size and phase of ${\cal{C}}_{8}$,
its correlation with ${\cal{C}}_{7}$,
and its interdependence with the branching ratio,
as well as its effects on the CP asymmetry 
have not been reported in detail. In
particular, given the post-LEP bound
on $\tan\beta\simgt 3.5$, it is
necessary to have a detailed knowledge of 
${\cal{C}}_{8}$ in this portion of the
SUSY parameter space. The
main goal of this work is to determine the
size and phase of the chromoelectric
coefficient ${\cal{C}}_{8}$ within the
CP--violating SUSY at beyond--the--leading--order (BLO)
precision in regions of the parameter space 
allowed by the existing $B\rightarrow X_s \gamma$
constraint.

The inclusive decay $B\rightarrow X_s \gamma$ is
well approximated (within at most $10\%$) by the
partonic decay $b\rightarrow s \gamma$ which is
described by the effective Hamiltonian 
\begin{eqnarray}
H_{eff}=- \frac{4 G_F}{\sqrt{2}} V^{\star}_{t s} V_{t b} \sum_{i=1}^{8} {\cal{C}}_{i}(\mu) {\cal{O}}_{i}(\mu) ~,
\end{eqnarray}
where $V$  is the Cabibbo-Kobayashi-Maskawa (CKM) matrix, and the operator basis ${\cal{O}}_{i=1,\cdots,8}$ is defined in \cite{buras,munz}. This 
OPE for the Hamiltonian separates the long--distance (the matrix elements of the 
local operators ${\cal{O}}_{i}$) and short--distance (associated Wilson coefficients ${\cal{C}}_{i}$)
at any scale $\mu\in (m_b, M_{W})$. Moreover, HQET approximates the inclusive rate by the partonic one 
(all terms being of the order ${\cal{O}}(\Lambda_{QCD}/m_b)$ and higher ones
are negligible) \cite{shifman}.

Evolution of the Wilson coefficients from $\mu=M_W$ down to $\mu=m_b$ level is
governed by the standard QCD RGEs:
\begin{eqnarray}
\label{c7mb}
{\cal{C}}_{2}(m_b)&=& 1/2 (\eta^{-12/23} + \eta^{6/23})\nonumber\\
{\cal{C}}_{7}(m_b)&=& {\cal{C}}_{7}(M_{W}) \eta^{16/23} + {\cal{C}}_{8}(M_{W})\frac{8}{3}(\eta^{14/23}-\eta^{16/23}) 
+ {\cal{C}}_{2}(M_{W}){\sum_{i=1}^{8} h_{i}\eta^{r_{i}}}\nonumber\\
{\cal{C}}_{8}(m_{b})&=& {\cal{C}}_{8}(M_{W})\eta^{14/23} + {\cal{C}}_{2}(M_{W}){\sum_{i=1}^{8} g_{i}\eta^{r_{i}}}
\end{eqnarray}
where $\eta=\alpha_s(M_W)/\alpha_s(m_b)$, and the numerical coefficients $h_{i}$ , $r_{i}$  
and $g_{i}$ are given in \cite{buras}. 

The initial values for the QCD RGEs,  ${\cal{C}}_{2,7,8}(M_W)$, depend on details
of the short--distance theory at $\mu\sim M_{W}$. In standard electroweak theory for instance,
one finds ${\cal{C}}_{2}(m_b)= 1.023$, ${\cal{C}}_{7}(m_b)= -0.312$ and ${\cal{C}}_{8}(m_b)= -0.148$
at BLO precision \cite{munz}. Consequently, the NP effects can appear  in various ways:
($i$) There can be observable deviations from these numbers with or without sign change, or
($ii$) the coefficients can take complex values. Each type of departure from the SM prediction implies
certain aspects of the weak--scale NP effects. For instance, for ($ii$), it is obvious that 
the NP brings new sources of CP violation, and necessarily, the CP asymmetry of the decay 
deviates from the SM prediction ($\simlt 1 \%$) \cite{neubert1,neubert2}.

In what follows, we will take SUSY with explicit CP violation as the NP candidate, and
concentrate on the results of \cite{olive} where the Wilson coefficients were 
computed at NLO precision for those threshold effects enhanced at large $\tan\beta$.
Within this framework ${\cal{C}}_{2}(M_W) = 1$ as in the SM, but the two dipole 
coefficients  ${\cal{C}}_{7,8}(M_W)$ are significantly modified compared to the LO results \cite{masiero}.
With the MFV scheme, only chargino--top squark, charged Higgs--top quark and 
$W$-boson--top quark loops give significant contributions  
\begin{eqnarray}
\label{c78}
{\cal{C}}_{7,8}(M_W)={\cal{C}}_{7,8}^{W} (M_W) + {\cal{C}}_{7,8}^{H}(M_W)   
+{\cal{C}}_{7,8}^{\chi} (M_W)
\end{eqnarray}
In CP--violating SUSY, at LO precison ${\cal{C}}_{7,8}^{W,H} (M_W)$ are 
always real; they do not contribute to CP--violating observables, 
such as the CP--asymmetry, in the decay. However, the chargino 
contribution is complex due to the $\mu$ parameter 
(having finite phase $\phi_{\mu}$) and the stop trilinear
coupling $A_{t}$ (having finite phase $\phi_{A}$) with 
approximate structure $e^{i(\phi_{\mu} + \phi_{A})}$ \cite{masiero}. 

With BLO precision, however, there are finite threshold 
corrections to each piece in (\ref{c78}) such that all three
contributions ${\cal{C}}_{7,8}^{W} (M_W)$, ${\cal{C}}_{7,8}^{H}(M_W)$
and ${\cal{C}}_{7,8}^{\chi} (M_W)$ are now complex. Moreover,
larger the $\tan\beta$ larger their imaginary parts, so that 
even naively one expects CP--violating effects to be enhanced at
large $\tan\beta$. Indeed, as reported in \cite{olive}, 
the BLO CP asymmetry is significantly larger than the 
LO one at sufficiently large $\tan\beta$. Although the
asymmetry remains $\simlt 8\%$ in both cases \cite{borzumati}, 
there occurs an enhanced sensitivity to $\tan\beta$ for 
the BLO case. 

In the following we will perform a numerical 
study of the electric and chromoelectric dipole
coefficients, and discuss their phenomenological
implications. In the numerical analysis, we
take: ($i$) the light stop $\tilde{t}_2$ and the charged Higgs $H^{\pm}$
are degenerate and weigh close to the weak scale,
$M_{\tilde{t}_2}=M_{H}=250\ {\rm GeV}$; ($ii$) the
sfermions of first two generations are heavy 
enough, so that one can  neglect their contribution to 
$B\rightarrow X_s \gamma$ (this is a viable
way of suppressing the one--loop EDMs \cite{olive});
($iii$) the SU(2) gaugino, the right--handed sbottom
and the heavy stop are heavy, $M_2=M_{\widetilde{t}_1}=\widetilde{m}_{b_R}= 
1\ {\rm TeV}$, and form the SUSY breaking scale; ($iv$) the stop
and sbottom trilinear couplings have the 
same phase, $\theta_{A_b}=\theta_{A_t}$, and 
the latter is degenerate with the $\mu$ parameter
$\left|\mu\right|=\left|A_b\right|=150\
{\rm GeV}$ (sbottom parameters are needed for 2-loop EDM calculations
\cite{2loop}); and finally ($v$) the light stop is 
dominantly right--handed to agree with the
electroweak precision data,  hence the stop mixing 
is to be sufficiently small ($\theta_{\widetilde{t}}=\pi/20$).
Moreover, we vary $\tan\beta$ from 10 to 50 and 
the phase $\phi_{A,\mu}$ from 0 to $\pi$ in forming the 
scatter plots. The parameter space mentioned here 
has been determined after trying several combinations,
and it is one those points yielding large CP--violation
in the system. One particularly notices that 
the lightest chargino is a Higgsino so that CP--violation
via chargino--stop contribution is enhanced. In minimal
supergravity, for instance, the lightest chargino 
is SU(2) gaugino and thus CP--violation is very much suppressed.
\begin{figure}[htb]
\vskip -3.0truein
\centering
\epsfxsize=6in
\leavevmode\epsffile{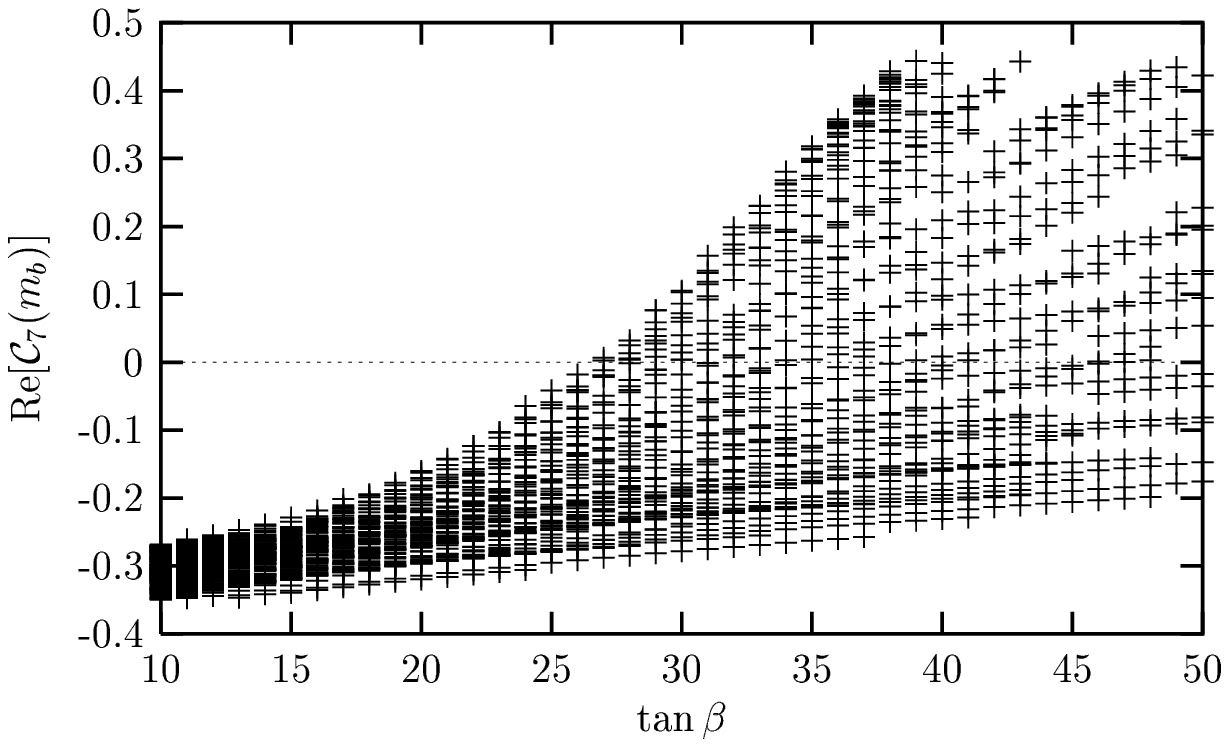}
\vskip -6.0truein
\centering
\epsfxsize=6in
\leavevmode\epsffile{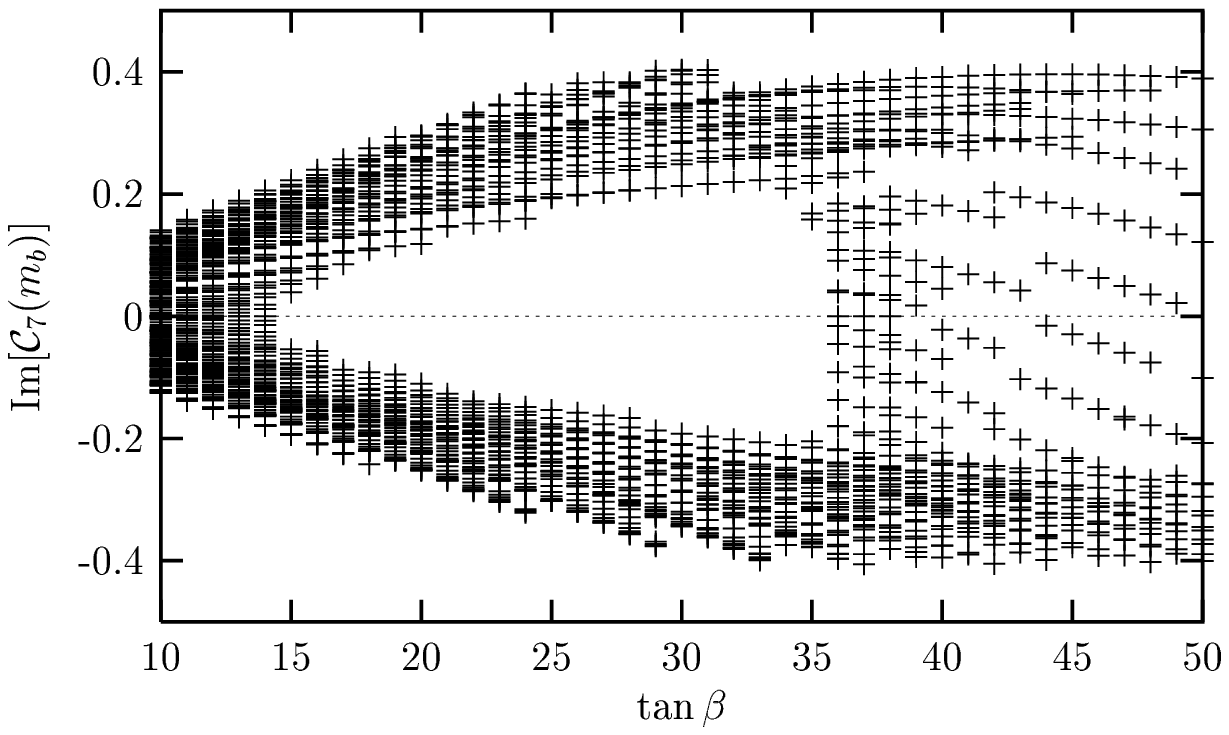}
\vskip -3.3truein
\caption[]{The $\tan\beta$ dependence of $\mbox{Re}[{\cal{C}}_{7}(m_b)]$ (upper window)
and $\mbox{Im}[{\cal{C}}_{7}(m_b)]$ (lower window), for the values of 
$\phi_{\mu}$ and $\phi_{A}$ varying  from 0 to $\pi$.}
\label{fig1}
\end{figure}

Depicted in Fig. 1 is the $\tan\beta$ dependence of
$\mbox{Re}[{\cal{C}}_{7}(m_b)]$ and $\mbox{Im}[{\cal{C}}_{7}(m_b)]$
for $10\leq \tan\beta\leq 50$ and $0\leq \phi_{\mu}, \phi_{A}\leq \pi$. 
When $\tan\beta\simgt 10$, $\mbox{Re}[{\cal{C}}_{7}(m_b)]$ is
close to the SM value and $\mbox{Im}[{\cal{C}}_{7}(m_b)]$ is 
evenly distributed around the origin, being consistent with 
zero. However, as $\tan\beta$ rises towards larger values,
so does $\mbox{Re}[{\cal{C}}_{7}(m_b)]$, crossing zero around
$\tan\beta\sim 27$. When $\tan\beta\simgt 27$, $\mbox{Re}[{\cal{C}}_{7}(m_b)]$
takes both negative and positive values, and it can be as large
as 0.4. The imaginary part of ${\cal{C}}_7$, however, rises
with increasing $\tan\beta$ in absolute magnitude for both negative and
positive directions 
in an approximately symmetric manner. For $\tan\beta\simgt 35$,
its distribution is levelled with a value swinging between 
-0.35 and +0.35 smoothly. Clearly, for large enough $\tan\beta$
there are regions of the parameter space where $\mbox{Im}[{\cal{C}}_{7}(m_b)]$
is not consistent with zero, signalling therefore a clear sign of
the NP effects. An approximate expression valid for
large $\tan\beta$ can be given as ${\cal{C}}_7\sim 0.4\ e^{\pm i\pi/4}$,
which is far away from the SM prediction in both size and phase.
However, there are regions of the parameter space where ${\cal{C}}_7$
is pure imaginary, is pure real or vanishes  exactly. 
\begin{figure}[htb]
\vskip -3.0truein
\centering
\epsfxsize=6in
\leavevmode\epsffile{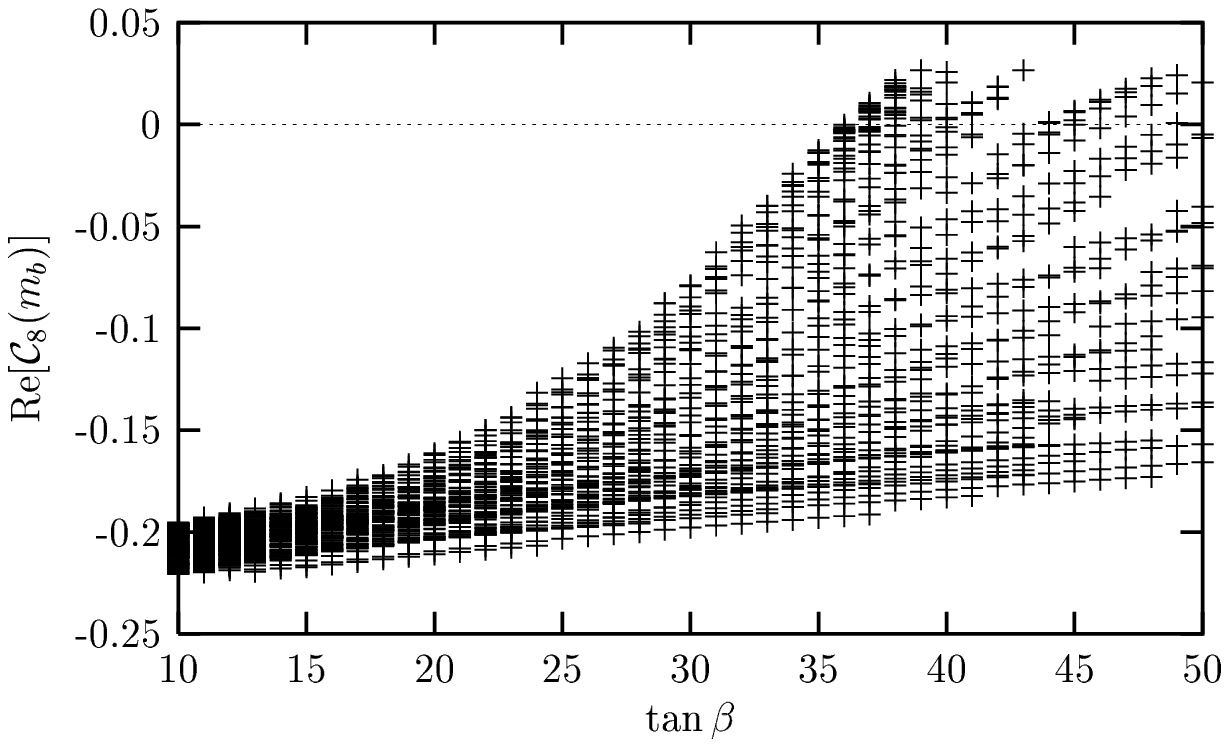}
\vskip -6.0truein
\centering
\epsfxsize=6in
\leavevmode\epsffile{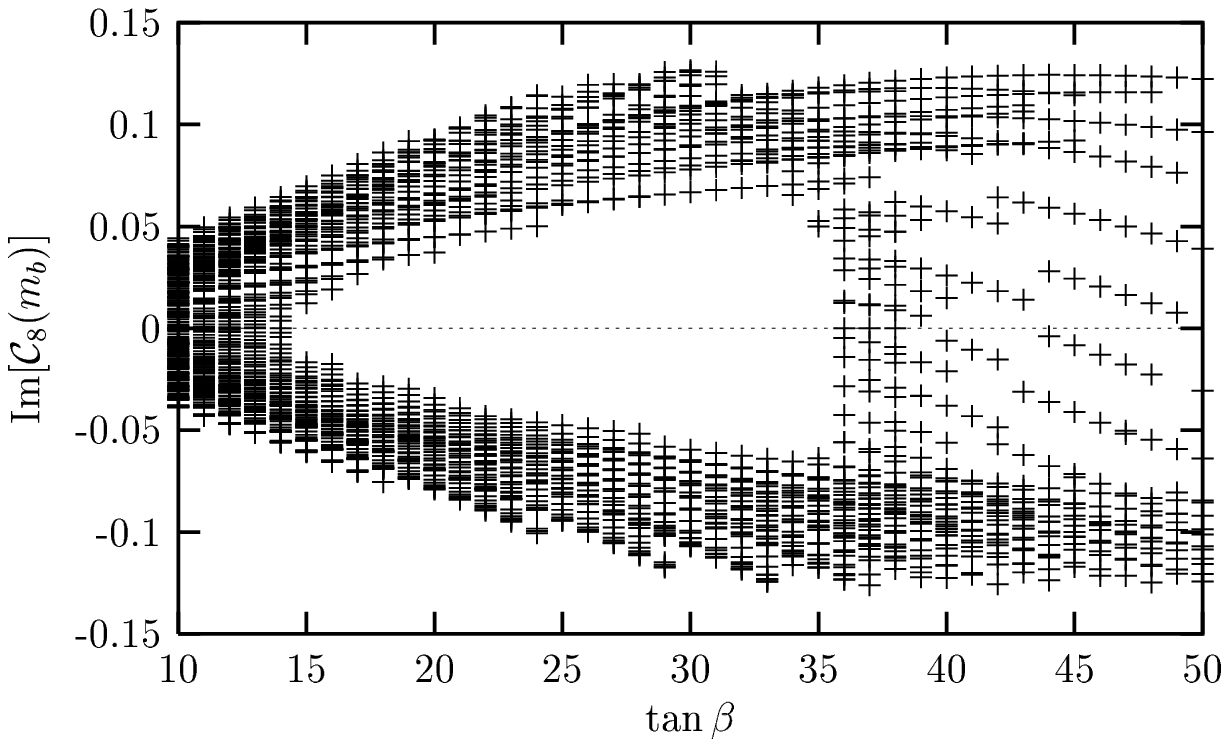}
\vskip -3.3truein
\caption[]{
 The $\tan\beta$ dependence of $\mbox{Re}[{\cal{C}}_{8}(m_b)]$ (upper window)
and $\mbox{Im}[{\cal{C}}_{8}(m_b)]$ (lower window), for the values of 
$\phi_{\mu}$ and $\phi_{A}$ varying  from 0 to $\pi$.}
\label{fig2}
\end{figure}

Similar to observations made for Fig. 1, one can discuss the
Wilson coefficient ${\cal{C}}_8$ using Fig. 2 where its
real (upper window) and imaginary (lower window) parts
are separately plotted against $\tan\beta$ when $\phi_{\mu,A}$
vary from 0 to $\pi$. One notices that, unlike ${\cal{C}}_7$,
for low $\tan\beta$, $\mbox{Re}[{\cal{C}}_{8}(m_b)]$ deviates
from its SM value. This stems from the fact that ${\cal{C}}_8(m_b)$
is directly proportional to ${\cal{C}}_8(M_W)$ (up to small
corrections proportional to ${\cal{C}}_2(M_W)$), and therefore
the NP effects at short distances are directly reflected to
the hadronic scale. As $\tan\beta$ rises to larger values,
both $\mbox{Re}[{\cal{C}}_{8}(m_b)]$ and $\mbox{Im}[{\cal{C}}_{8}(m_b)]$
gradually increase  where the former crosses zero around $\tan\beta\sim 35$.
In large $\tan\beta$ regime, $\mbox{Re}[{\cal{C}}_{8}(m_b)]$
can  take positive values only in a small portion
of the parameter space whereas $\mbox{Im}[{\cal{C}}_{8}(m_b)]$ swings
between -0.1 and 0.1 almost evenly. One notices that, as in ${\cal{C}}_7$,
in certain regions of the parameter space ${\cal{C}}_8(M_W)$ can be
pure real, pure imaginary or just vanish. Those points where
both  ${\cal{C}}_7$ and ${\cal{C}}_8$ vanish are particularly
interesting, as 
in this case one has to saturate the experimental bounds  on 
$B\rightarrow X_s \gamma$ via the chirality--flipped Wilson 
coefficients, implying large contributions due to 
the gluino exchange \cite{everett}. As we are working in the
MFV scheme, such effects are obviously beyond the scope of our 
discussion.

Depicted in Fig. 3 is the dependence of $\mbox{Im}[{\cal{C}}_{8}(m_b)]$
on $\mbox{Re}[{\cal{C}}_{7}(m_b)]$. Here the size and shape
of the ellipsoidal region is determined by the accuracy of 
the experimental results on $\mbox{BR}(B\rightarrow X_s \gamma)$.
The main NP effect occurs in pushing $\mbox{Re}[{\cal{C}}_{7}(m_b)]$
to larger values compared to the SM prediction instead of 
smaller ones. The small region around $\mbox{Re}[{\cal{C}}_{7}(m_b)]\sim -0.3$
corresponds to the SM validity domain where imaginary 
parts of both coefficients remain around zero. 
\begin{figure}[htb]
\vskip -3.0truein
\centering
\epsfxsize=6in
\leavevmode\epsffile{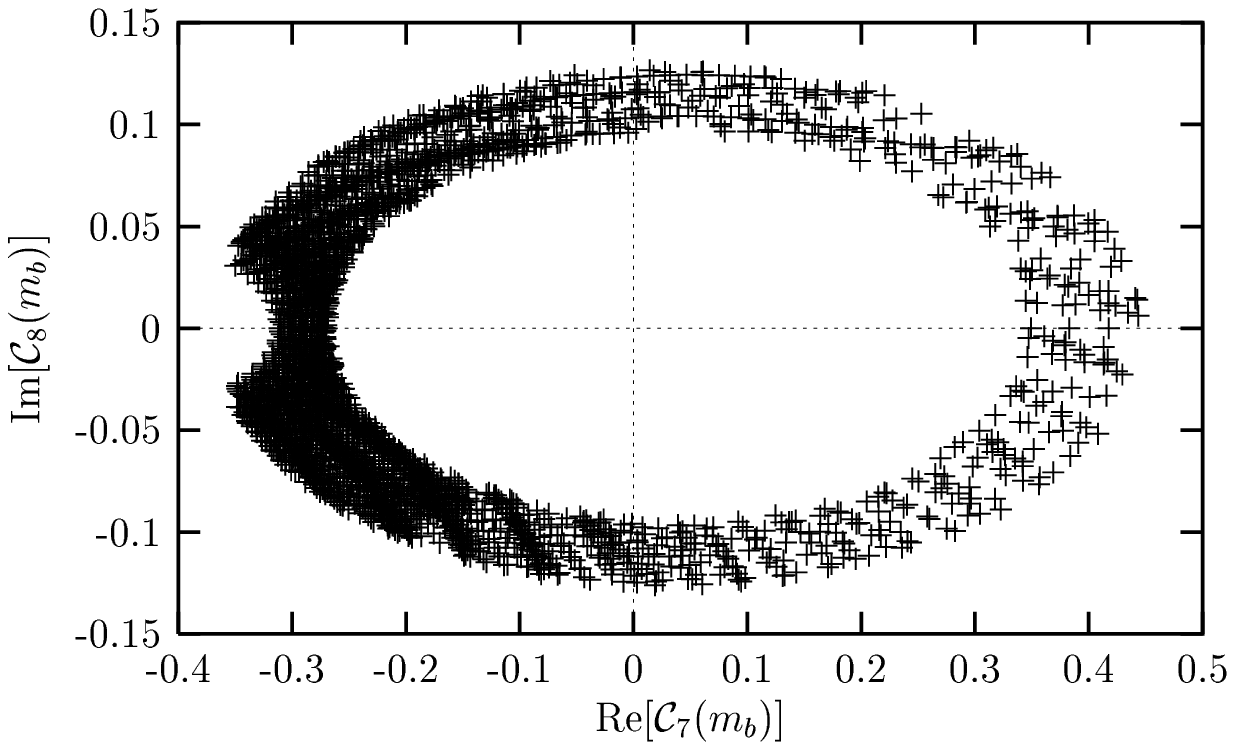}
\vskip -3.3truein
\caption[]{ Dependence of $\mbox{Im}[{\cal{C}}_{8}(m_b)]$ on $\mbox{Re}[{\cal{C}}_{7}(m_b)]$.
The width of the ellipsoid corresponds to present experimental accuracy 
for $\mbox{BR}(B\rightarrow X_s \gamma)$ .}
\label{fig3}
\end{figure}

Fig. 4 illustrates the dependence of $\mbox{A}_{\small \mbox{CP}}(B\rightarrow X_s \gamma)$
on $\mbox{Re}[{\cal{C}}_{8}(m_b)]$ (upper window), and on $\mbox{Im}[{\cal{C}}_{8}(m_b)]$ (lower
window). As we can see from the  upper window, the asymmetry takes the
largest value $\pm 8\%$ when $\mbox{Re}[{\cal{C}}_{8}(m_b)]$ is in the
range of the SM prediction for ${\cal{C}}_{8}(m_b)$.
One notes that in a more general model CP asymmetry can be as large as 
$ 20\%$\cite{neubert2}. Moreover, in certain small corners of the SUSY parameter
space the CP asymmetry can be enhanced by a factor of 2 \cite{baek}. 
When $\mbox{Re}[{\cal{C}}_{8}(m_b)]$ 
takes larger or smaller values than the SM prediction, the asymmetry gradually
drops to the corresponding SM prediction $\sim 1\%$. The dependence of 
the CP asymmetry on $\mbox{Im}[{\cal{C}}_{8}(m_b)]$, however, shows that 
the maximal values are attained when $\mbox{Im}[{\cal{C}}_{8}(m_b)]\sim \pm
0.1$, which is far away from the SM predicton. In conclusion, the asymmetry is
roughly an order of magnitude larger than what is expected in the SM,
and this happens when $\mbox{Re}[{\cal{C}}_{8}(m_b)]$ remains close
to the SM prediction, and $\mbox{Im}[{\cal{C}}_{8}(m_b)]$ is large
and has a similar size as the real part.  
\begin{figure}[htb]
\vskip -3.0truein
\centering
\epsfxsize=6in
\leavevmode\epsffile{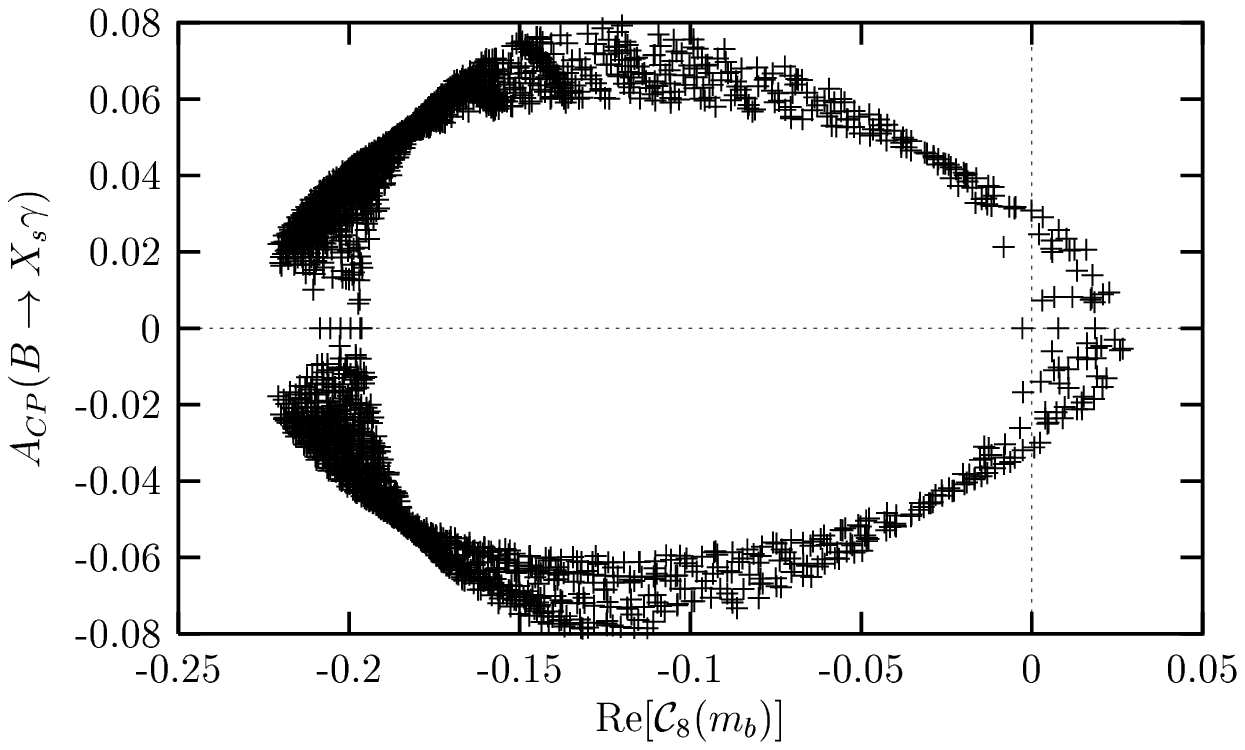}
\vskip -6truein
\centering
\epsfxsize=6in
\leavevmode\epsffile{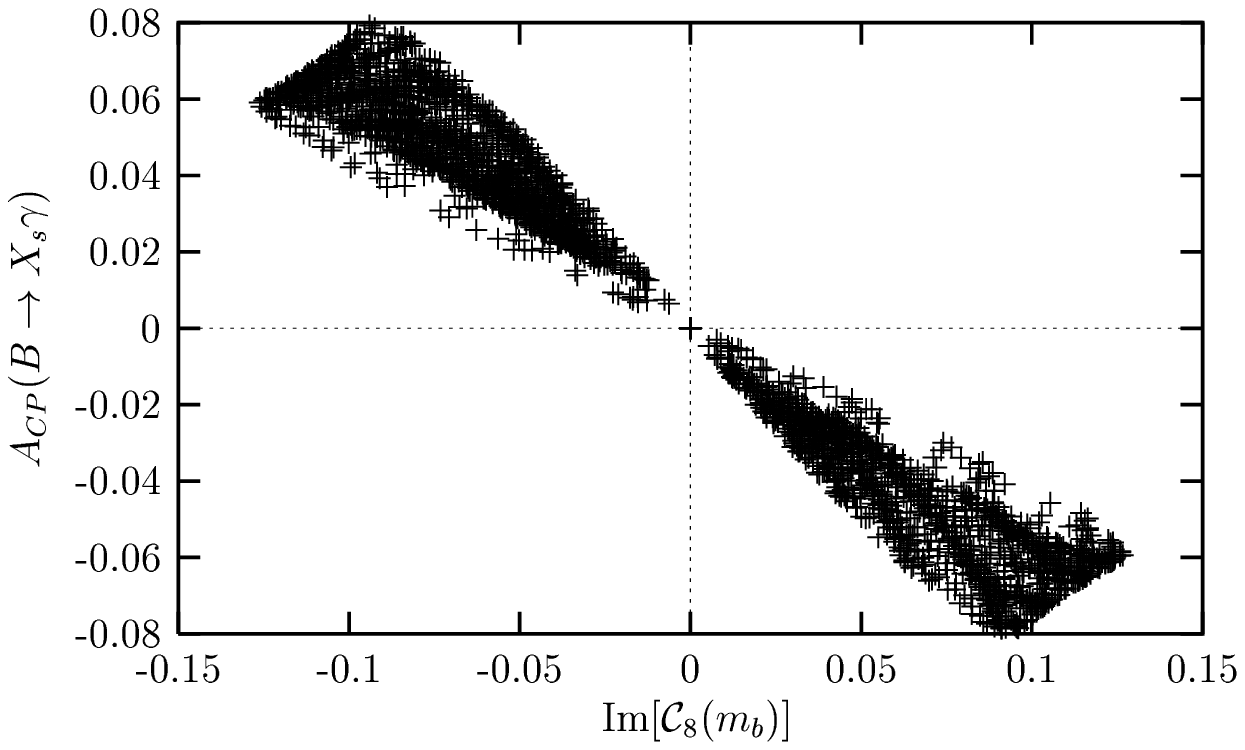}
\vskip -3.3truein
\caption[]{
The dependence of $\mbox{A}_{\small \mbox{CP}}(B\rightarrow X_s \gamma)$
on $\mbox{Re}[{\cal{C}}_{8}(m_b)]$ (upper window) and on
$\mbox{Im}[{\cal{C}}_{8}(m_b)]$ (lower window).}
\label{fig4}
\end{figure}

A closer comparative look at the figures suggests that the CP
asymmetry, which will be measured in near--future $B$ factories
with increasing precision, is maximal when $|{\cal{C}}_{8}(m_b)|\sim
|{\cal{C}}_{7}(m_b)|$. This results confirms earlier predictions \cite{wolf}
where it was already shown that $\mbox{A}_{\small \mbox{CP}}(B\rightarrow X_s \gamma)
\sim 10 \%\ |{\cal{C}}_{8}(m_b)/{\cal{C}}_{7}(m_b)|$.

Even after including the NLL corrections to the semileptonic
inclusive $B$ decay $B\rightarrow X e \nu$, the theoretical
predictons for the branching ratio and the charm multiplicity
turn out to be larger than the experimental result \cite{bagan,besmer}.
Therefore, it is conceivable that possible enhancements in ${\cal{C}}_{8}$
can account for the existing discrepancy between the theory and the 
experiment. As the numerical analyses above show, it is possible to
significantly shift this coefficient in both real and imaginary directions
compared to the SM prediction. Therefore, the experimental 
result on $B\rightarrow X_s \gamma$ allows for large deviations
in chromoelectric as well as electric coefficients at large 
values of $\tan\beta$ where the SUSY threshold corrections are important.
Concerning the semileptonic $B$ decays, the exclusive channel 
$B\rightarrow X_s \ell^+\ell^-$ is another example having both 
theoretical and experimental importance. This follows from the
fact that the forward--backward asymmetry $\mbox{A}_{\small \mbox{FB}}$
of this decay vanishes at a specific value of the dilepton invariant
mass \cite{bsll} in a hadronically clean way. This zero of
$\mbox{A}_{\small \mbox{FB}}$ occurs at the point 
$m_{\ell\ell}^{2}\sim - m_{b}^2 \mbox{Re}\left[{\cal{C}}_{7}(m_b)/{\cal{C}}_{9}(m_b)\right]$
where ${\cal{C}}_{9}(m_b)$ is the four--fermion operator coefficient 
(not computed here). The important point is that the position
of the zero shifts in accord with the NP contributions
to the Wilson coefficients. Suppose that ${\cal{C}}_{7}(m_b)$ alone
is shifted in complex direction, then it is clear that the critical
value of the dilepton mass $m_{\ell\ell}^{2}$ is shifted back and
forth, depending on the parameter values \cite{bsll}.

There are other processes  where the radiatively corrected Wilson
coefficients play  an important role. As an application,  
for instance, we analyze the bottom baryon radiative decay, $\Lambda_b \rightarrow \Lambda
\gamma $, which is again dominated by the mechanism $b \rightarrow
s \gamma$.
We will  particularly concentrate on  the dependence of the  braching ratio
on  the Wilson coefficients at the weak scale.

The decay rate of $ \Lambda_b \rightarrow \Lambda \gamma$ 
has been computed in \cite{cheng}, and  its expression is given by: 
\begin{eqnarray}
\label{rate}
\Gamma (\Lambda_b \rightarrow \Lambda \gamma )&=&  \frac{1}{8
\pi}\Bigg(
\frac{m_{{\Lambda_{b}}}^{2}-m_{\Lambda}^{2}}{m_{{\Lambda_{b}}}}\Bigg)^3
\Bigg(|a|^2+|b|^2 \Bigg)~,
\end{eqnarray}
with  
\begin{eqnarray}
a&=&\frac{G_{F}}{\sqrt{2}}\frac{e}{8 \pi^2} 2 C_{7}(m_b) m_{b} V_{tb}
V_{ts}^{*}
\Bigg(f_{1}^{\Lambda\Lambda_b}(0)-f_{2}^{\Lambda\Lambda_b}(0)\Bigg)~,\nonumber\\
b&=&\frac{G_{F}}{\sqrt{2}}\frac{e}{8 \pi^2} 2 C_{7}(m_b) m_{b} V_{tb}
V_{ts}^{*}
\Bigg(g_{1}^{\Lambda\Lambda_b}(0)+
\Big( \frac{m_{{\Lambda_{b}}}-m_{\Lambda}}{m_{{\Lambda_{b}}}+m_{\Lambda}}\Big)
g_{2}^{\Lambda\Lambda_b}(0)\Bigg)~.
\end{eqnarray}
The form factors in $a$ and $b$ coefficients, depend on the momentum transfer
as follows \cite{cheng, cheng2}:
\begin{eqnarray}
f_{i}(q^{2})=f_{i}(q_{m}^{2})\Bigg (
\frac{1-q_{m}^{2}/m_{V}^{2}}{1-q^{2}/m_{V}^{2}}\Bigg)^{n}~,\,\,\,\,\,\,
g_{i}(q^{2})=g_{i}(q_{m}^{2})\Bigg (
\frac{1-q_{m}^{2}/m_{A}^{2}}{1-q^{2}/m_{A}^{2}}\Bigg)^{n}~,
\end{eqnarray}
where $q_m^{2}=(m_{{\Lambda_{b}}}-m_{\Lambda})^{2}$ and, $q^{2}=p_{{\Lambda_{b}}}-p_{\Lambda}$,
with n=1 and n=2 representing  the monopole and dipole 
contributions. Here,  
$m_V$ and  $m_A$ are the pole masses of the vector and axial vector
mesons, respectively. 

In the numerical analysis,
following \cite{cheng, cheng2}, we let \,
$f_{1}^{\Lambda\Lambda_b}(q_m^2)=g_{1}^{\Lambda\Lambda_b}(q_m^2)=0.64$,
\,\, $g_{2}^{\Lambda\Lambda_b}(q_m^2)=-0.10$, and
$f_{2}^{\Lambda\Lambda_b}(q_m^2)=-0.31$ for the values of the heavy light form factors.
Additionaly, we take $m_b(m_b)=4.25 \,\mbox{GeV}$,
$m_{{\Lambda_{b}}}=5.624 \,\mbox{GeV}$, $m_V=5.42 \,\mbox{GeV}$, $m_A=5.86
\,\mbox{GeV}$, and $\tau (\Lambda_b)=1.23 \times 10^{-12} \mbox{sec}$. 
In forming the scatter plots,
we vary $\tan\beta$  from 10 to 50 and
the phase  $\phi_{A,\mu}$ from 0 to $\pi$. 
\begin{figure}[htb]
\vskip -3.0truein
\centering
\epsfxsize=6in
\leavevmode\epsffile{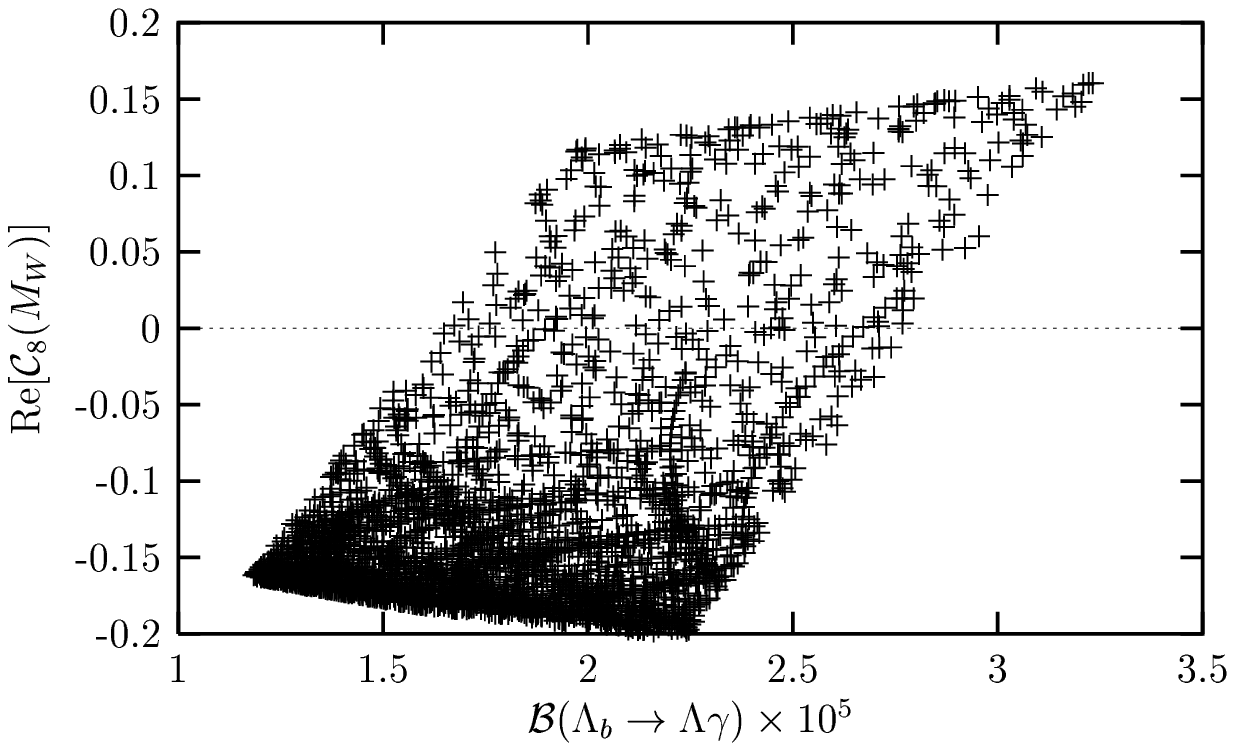}
\vskip -6truein
\centering
\epsfxsize=6in
\leavevmode\epsffile{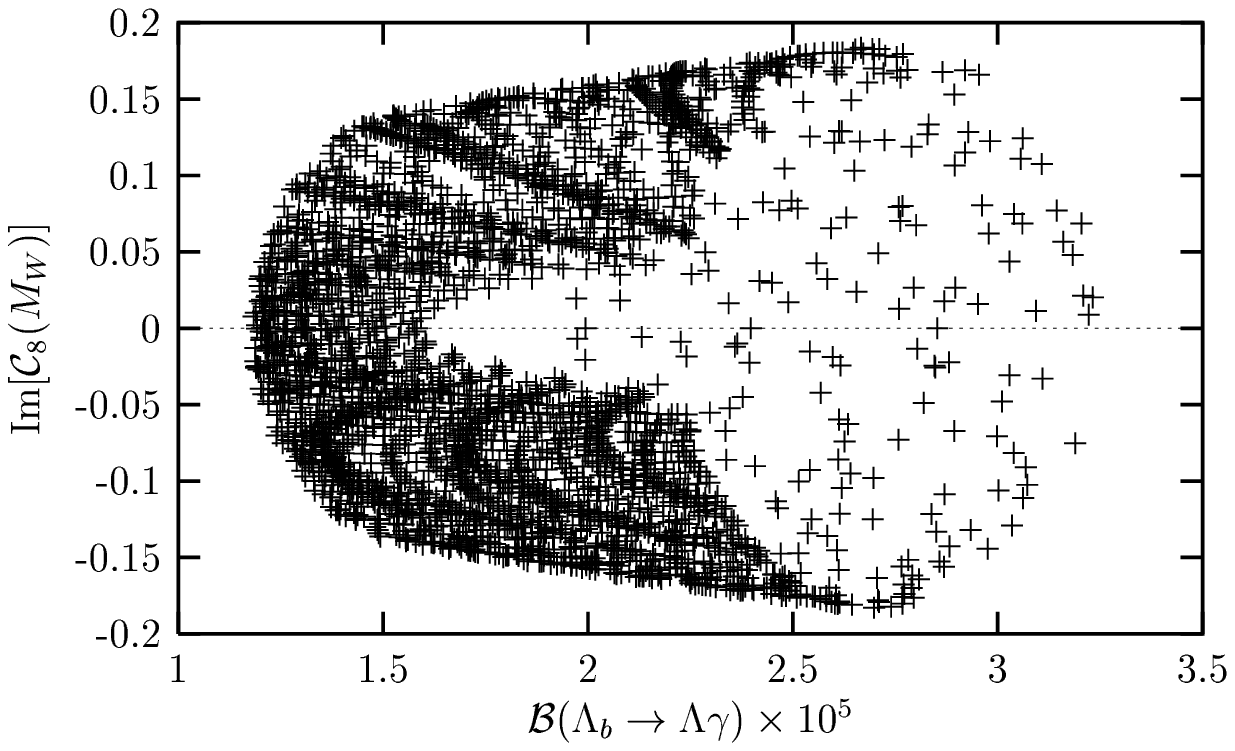}
\vskip -3.3truein
\caption[]{The dependence of
${\cal B}(\Lambda_b \rightarrow \Lambda \gamma)$
on $\mbox{Re}[{\cal{C}}_{8}(M_W)]$ (upper window) and on
$\mbox{Im}[{\cal{C}}_{8}(M_W)]$ (lower window), for the values of 
$\phi_{\mu}$ and $\phi_{A}$ varying  from 0 to $\pi$.}
\label{fig5}
\end{figure}

Depicted in Fig. 5 is the dependence of the branching ratio of $\Lambda_b
\rightarrow \Lambda \gamma$ (${\cal B}(\Lambda_b \rightarrow \Lambda \gamma )$) on
$\mbox{Re}[{\cal{C}}_{8}(M_W)]$ (upper window) and $\mbox{Im}[{\cal{C}}_{7}(M_W)]$
(lower window), for $10\leq \tan\beta\leq 50$ and $0\leq \phi_{\mu}, \phi_{A}\leq \pi$,
where only the monopole $q^2$ dependence for baryon form factors is considered (n=1).
As is seen from  both windows, $\mbox{Re}[{\cal{C}}_{8}(M_W)]$
takes negative values up to ${\cal B}(\Lambda_b \rightarrow \Lambda \gamma )\sim
1.7\times10^{-5}$, whereas $\mbox{Im}[{\cal{C}}_{8}(M_W)]$
is evenly distributed around the origin in this interval. 
$\mbox{Re}[{\cal{C}}_{8}(M_W)]$ starts to take both positive and negative
values, when  ${\cal B}(\Lambda_b \rightarrow
\Lambda \gamma )\simgt 1.7\times10^{-5}$. It can take  only positive values
for ${\cal B}(\Lambda_b \rightarrow \Lambda \gamma )\simgt 2.7\times 10^{-5}$,  
and can be as large as $\sim 0.2$. 
On the other hand,  
for  ${\cal B}(\Lambda_b \rightarrow
\Lambda \gamma)\simgt 1.7\times10^{-5}$,  
$\mbox{Im}[{\cal{C}}_{8}(M_W)]$
increases in both positive and negative directions in a symmetric manner. 
Clearly, 
there are  certain regions of the parameter space where ${\cal{C}}_8(M_W)$ can be
pure real, pure imaginary as well as it  just vanishes.  

In the recent work \cite{cheng}, 
it has been shown  that
${\cal B}(\Lambda_b \rightarrow
\Lambda \gamma )$ has a magnitude of $1.9 \times 10^{-5}$,
when only the monopole $q^2$ dependence of the  baryon form factors  
is considered (n=1).
This  decay   has also been considered  in \cite{ cheng3}
with the predicted branching ratios in the range  of $(1.2-1.9)\times 10^{-5}$.
As  one can see from both windows of  Fig. 5,  ${\cal B}(\Lambda_b \rightarrow
\Lambda \gamma)$ varies  in the range of $(1 -3.2)\times 10^{-5}$, 
and
the maximal values of ${\cal B}(\Lambda_b \rightarrow
\Lambda \gamma )$ are attained  when
$\mbox{Re}[{\cal{C}}_{8}(M_W)] \sim 0.2 $,
and  $\mbox{Im}[{\cal{C}}_{8}(M_W)] \sim \pm 0.2$,
which are away from the S.M prediction for ${\cal{C}}_{8}(M_W)$. 
A closer comparative look at the figure suggests that
when ${\cal{C}}_8(M_W)$ is in the range of the S.M prediction,
for instance  ${\cal{C}}_8(M_W)=-0.086$ \cite{buras},
${\cal B}(\Lambda_b \rightarrow
\Lambda \gamma )$ drops nearly to  $\sim 1.9\times 10^{-5}$, which is  consistent with the   
results of S.M\cite{cheng}. However, one notes that at this particular value
of the branching ratio, there are other solutions in the parameter space where
${\cal{C}}_8(M_W)$ can deviate from its S.M value  
not only in sign, but also in magnitude, as it  can be complex.  

We would like to note that  the analysis of \cite{cheng} has been  carried out in the
context of the S.M, and $C_7(m_b)$ has a fixed value $(C_7(m_b)= -0.312)$. However,
in our work, $C_7$ and  $C_8$ are complex, and
as we have shown in Figs. 1 and 2, they vary in a wider range.
For instance,  approximate expressions of  $C_{7(8)}$  valid for
large $\tan\beta$ can be given as ${\cal{C}}_7\sim 0.4\ e^{\pm i\pi/4}$,
and  ${\cal{C}}_8\sim 0.1\ e^{\pm i\pi/4}$,  
which are far away from the S.M predictions in both size and phase.
Therefore, having a wide range of parameter space, the model under concern gives new
allowed regions. It is natural that there are regions in the parameter space where our
prediction for  ${\cal B}(\Lambda_b \rightarrow
\Lambda \gamma )$ agrees with that of S.M.
\begin{figure}[htb]
\vskip -3.0truein
\centering
\epsfxsize=6in
\leavevmode\epsffile{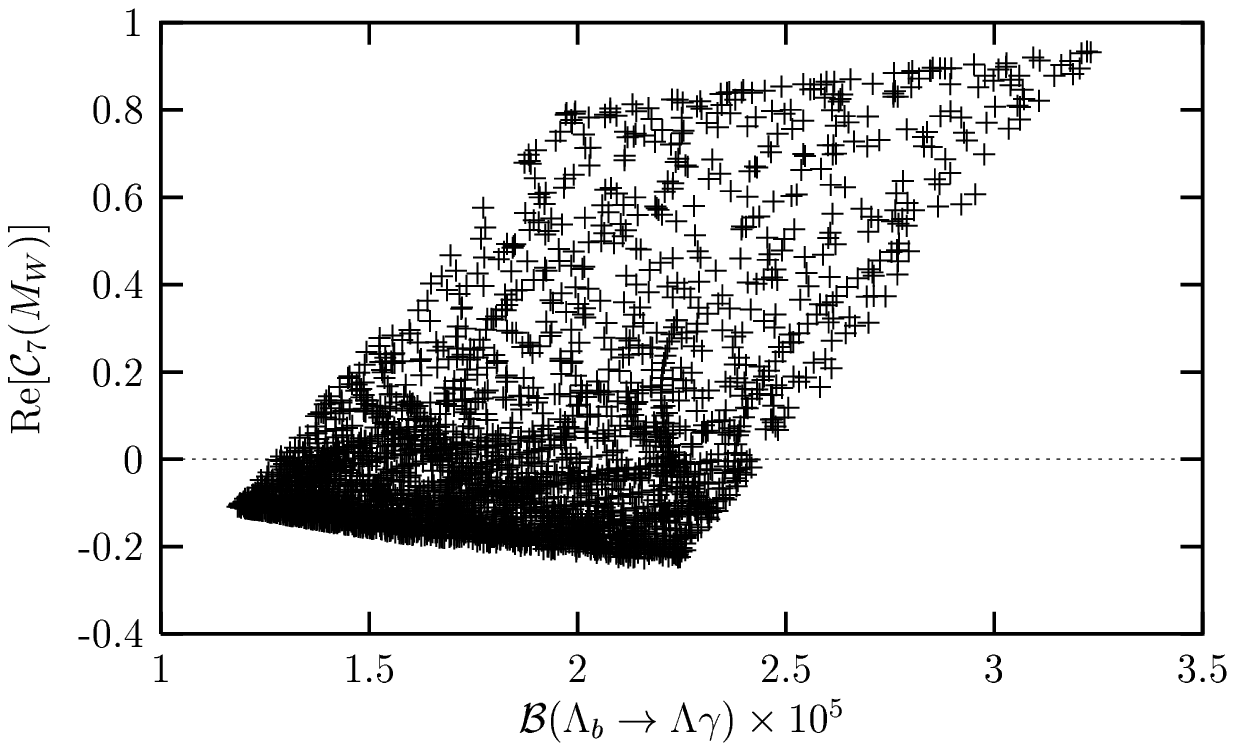}
\vskip -6truein
\centering
\epsfxsize=6in
\leavevmode\epsffile{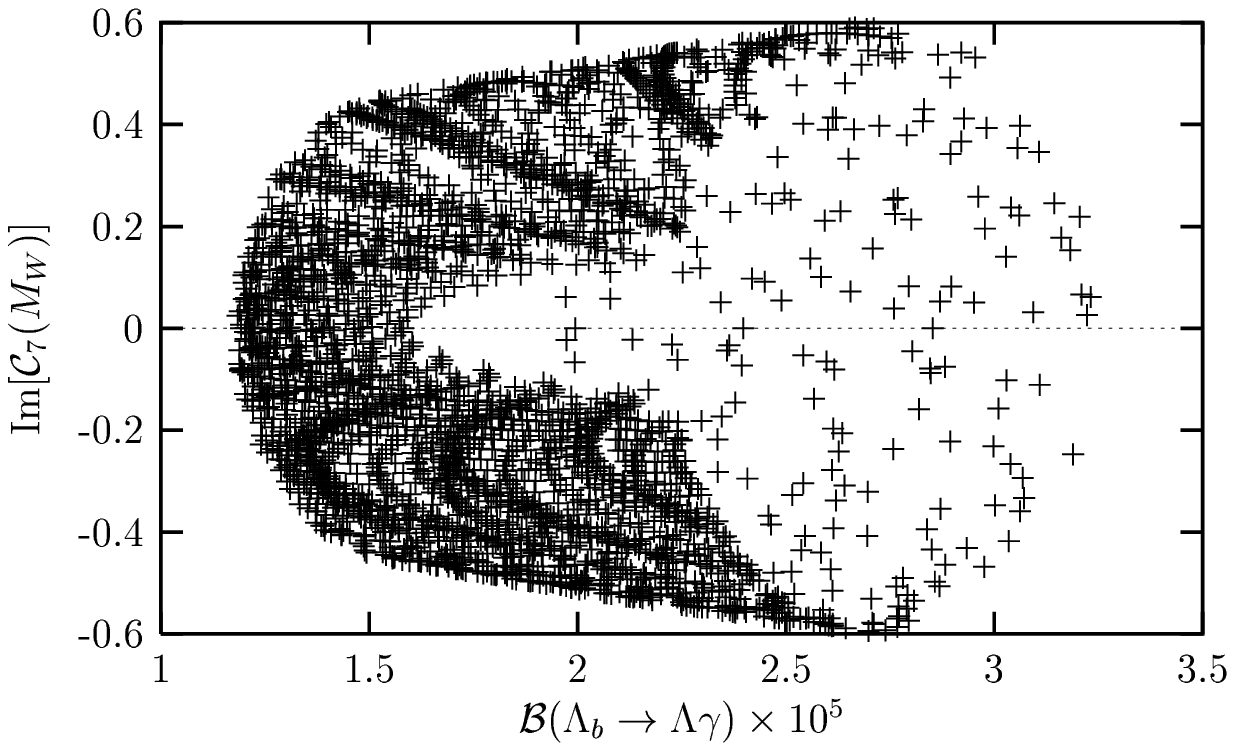}
\vskip -3.3truein
\caption[]{The dependence of 
${\cal B}(\Lambda_b \rightarrow \Lambda \gamma)$ on $\mbox{Re}[{\cal{C}}_{7}(M_W)]$ (upper window) and on
$\mbox{Im}[{\cal{C}}_{7}(M_W)]$ (lower window) , for the values of 
$\phi_{\mu}$ and $\phi_{A}$ varying  from 0 to $\pi$.}
\label{fig6}
\end{figure}

Similar to the observations  made for Fig. 5,
one can discuss the
Wilson coefficient ${\cal{C}}_7$ using Fig. 6, where its
real (upper window) and imaginary (lower window) parts
are separately plotted against ${\cal B}(\Lambda_b \rightarrow \Lambda \gamma)$,
when  $\phi_{\mu,A}$ vary from 0 to $\pi$, and
only monopole contribution is considered (n=1). 
As  ${\cal B}(\Lambda_b \rightarrow
\Lambda \gamma)$ rises to larger values, both  $\mbox{Re}[{\cal{C}}_{7}(M_W)]$ and
$\mbox{Im}[{\cal{C}}_{7}(M_W)]$ gradually increase, where the former
crosses zero around ${\cal B}(\Lambda_b \rightarrow \Lambda \gamma)\sim 1.3 \times
10^{-5}$. $\mbox{Re}[{\cal{C}}_{7}(M_W)]$ starts to take positive values only
when  ${\cal B}(\Lambda_b \rightarrow \Lambda \gamma)\simgt 2.4 \times
10^{-5}$, whereas $\mbox{Im}[{\cal{C}}_{7}(M_W)]$
swings between $\pm 0.6$. 
Like ${\cal{C}}_8$, there are observable deviations from the S.M
prediction with or without sign change, and it can take  complex values
as well.
However, a comparative glance at the figure shows that,  
when ${\cal{C}}_{7}(M_W)$
is in the range of the SM prediction,
for instance ${\cal{C}}_7(M_W)=-0.161$\cite{buras},  
${\cal B}(\Lambda_b \rightarrow
\Lambda \gamma )\sim 1.3\times 10^{-5}$, which is  close to 
the S.M value \cite{cheng, cheng3}.
\begin{figure}[htb]
\vskip -3.0truein
\centering
\epsfxsize=6in
\leavevmode\epsffile{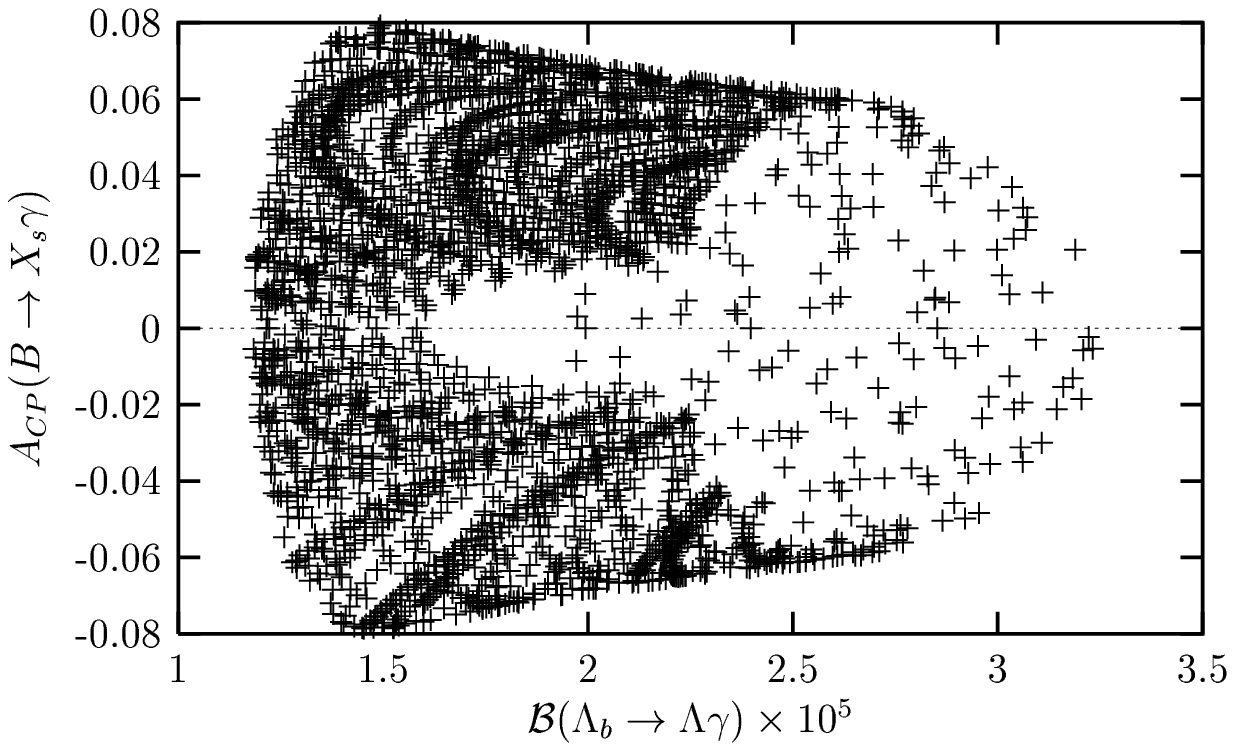}
\vskip -6truein
\centering
\epsfxsize=6in
\leavevmode\epsffile{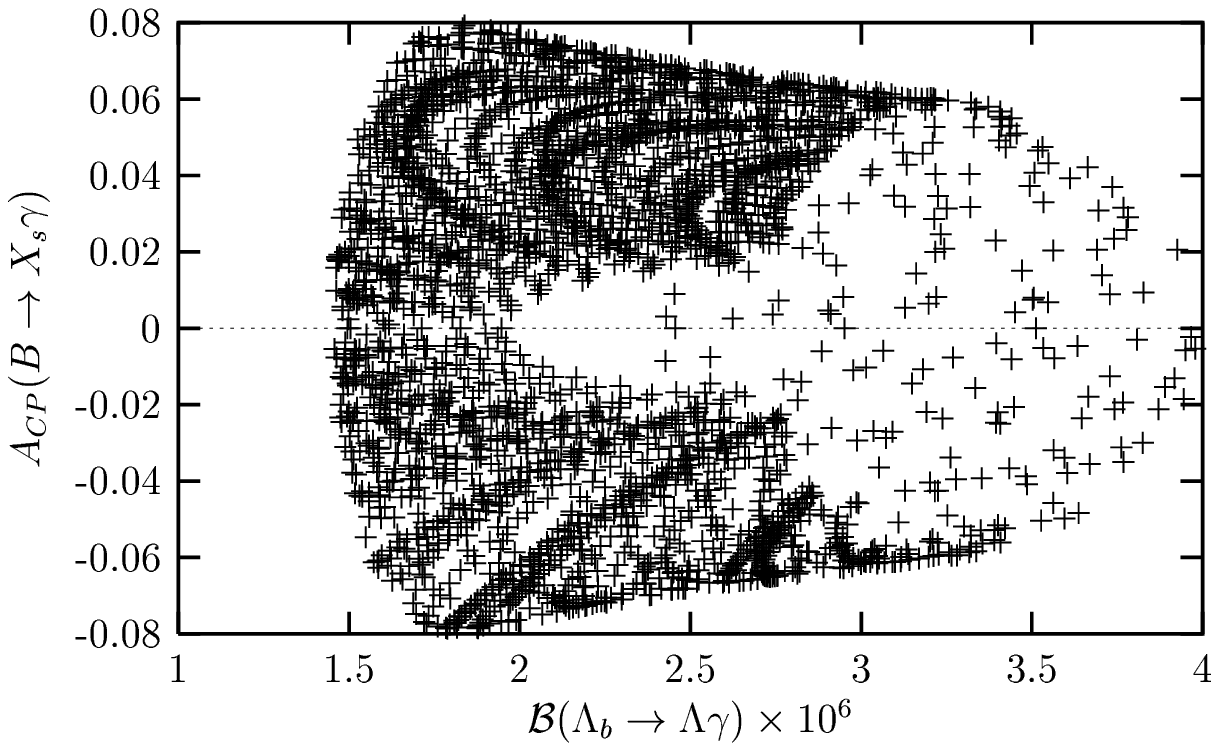}
\vskip -3.3truein
\caption[]{The dependence of ${\cal B}(\Lambda_b \rightarrow \Lambda \gamma)$
on $\mbox{A}_{\small \mbox{CP}}(B\rightarrow X_s \gamma)$  for n=1 (upper window) and 
for n=2  (lower window)}
\label{fig7}
\end{figure}

Depicted in Fig. 7 is the dependence of  ${\cal B}(\Lambda_b \rightarrow
\Lambda \gamma)$
on  $\mbox{A}_{\small \mbox{CP}}(B\rightarrow X_s \gamma)$, when
only the monopole  (upper window)
and  the dipole  (lower window) $q^2$ dependences are considered.
As is noticed from the upper and lower windows   ${\cal B}(\Lambda_b
\rightarrow \Lambda \gamma)$
is of the orders of $10^{-5}$ and $10^{-6}$, for the monopole (n=1)  and
dipole (n=2) contributions, respectively. 
In both cases  ${\cal B}(\Lambda_b \rightarrow
\Lambda \gamma)$ behave similarly. 
However, it decreases by an order of magnitude in the dipole case,  
and the maximal values of  $\mbox{A}_{\small \mbox{CP}}(B\rightarrow X_s \gamma)$
are obtained when  ${\cal B}(\Lambda_b \rightarrow
\Lambda \gamma)\sim 1.5 \times 10^{-5} (1.9 \times 10^{-6})$ for n=1 (2).
In the recent work of \cite{cheng}, it has been shown that  ${\cal B}(\Lambda_b
\rightarrow\Lambda \gamma)=2.3 \times 10^{-6}$ for n=2.
As we can see from the  upper and lower windows, the asymmetry  nearly
takes the  largest value, when  ${\cal B}(\Lambda_b \rightarrow
\Lambda\gamma)$ is in the range of the SM prediction\cite{cheng, cheng4, singer}.
When ${\cal B}(\Lambda_b \rightarrow
\Lambda\gamma)$  takes larger  values than the SM prediction, the asymmetry gradually
drops to the corresponding SM prediction $\sim 1\%$. 

In conclusion, we have computed the dipole coefficients
${\cal{C}}_{7,8}$ in SUSY with explicit CP violation with
special emphasis on large values of $\tan\beta$. We have shown
that the present experimental bounds on $B\rightarrow X_s \gamma$
allows for large deviations in the Wilson coefficients (with
respect to the SM prediction) in both real and imaginary directions.
The CP asymmetry in the decay is enhanced by an order of
magnitude, thanks to especially the SUSY threshold corrections.
The allowed deviations from the SM values can account
for (being a plausable hypothesis) the discrepancy between
the experiment and theory for the semileptonic
$B$ decays. As an illustration, we have discussed
$\Lambda_b \rightarrow \Lambda \gamma$
decay.

\noindent
M. B would like to thank the Scientific and Technical Research
Council of Turkey (T\"{U}B{\.I}TAK) for partial support under the project,
No:TBAG2002(100T108).

\end{document}